\documentclass[10pt,letterpaper]{article}

\usepackage[margin=1in]{geometry}
\usepackage{times}
\usepackage[dvipsnames]{xcolor}
\usepackage{graphicx}
\usepackage{amsmath,amssymb}
\usepackage{booktabs}
\usepackage[numbers,sort&compress]{natbib}
\usepackage{caption}
\usepackage{subcaption}
\usepackage{enumitem}
\usepackage{url}

\usepackage{multirow}
\usepackage{algorithm}
\usepackage{algpseudocode}

\newcommand{\TODO}[1]{}

\usepackage{microtype}

\definecolor{cvprblue}{rgb}{0.21,0.49,0.74}
\usepackage[breaklinks,colorlinks,allcolors=cvprblue]{hyperref}

\title{Mitigating Compiler Fusion-Induced Power Bursts in Mobile NPU Inference as the Battery Depletes}

\author{
Ryoga Yuzawa$^{1,2}$\thanks{Visiting researcher at University of California, Berkeley.} \quad Masayoshi Tomizuka$^{1}$\\
$^{1}$UC Berkeley \quad $^{2}$Sony
}
\date{}
\hypersetup{
    pdftitle={Mitigating Compiler Fusion-Induced Power Bursts in Mobile NPU Inference as the Battery Depletes},
    pdfauthor={Ryoga Yuzawa and Masayoshi Tomizuka},
    pdfsubject={arXiv preprint}
}

\begin{document}
\maketitle
\begin{center}
\bfseries Abstract
\end{center}
\noindent
Mobile devices increasingly rely on real-time NPU inference for
camera and perception workloads.
Under low-voltage conditions, however, a single inference can induce
an instantaneous voltage droop on the power-delivery network (PDN),
forcing the PMIC to invoke dynamic voltage and frequency scaling
(DVFS) and inflating latency.

This paper presents a measurement study of that effect on a commercial
smartphone.
We show that aggressive operator fusion in a mobile NPU compiler can
create monolithic superlayers whose concentrated execution
produces large peak-current bursts.
These bursts shift the DVFS-onset voltage upward and reduce the
low-voltage operating margin.

We further evaluate a practical black-box mitigation: a
measurement-guided pre-compilation graph rewrite whose deployed form
inserts barriers at selected PAPR hot spots to block harmful
superlayer merging in the vendor NPU compiler.
On Snapdragon~8~Gen~3 with MobileNetV4 (768$\times$768, ImageNet-1k),
this reduces peak current from 3.12\,A to 1.94\,A with 3.76\,\%
latency overhead, preserves stable latency deeper into the
low-voltage regime, and shifts the inferred DVFS margin by
approximately 173\,mV.
 \section{Introduction}
\label{sec:intro}

Deep neural networks (DNNs) are increasingly deployed on mobile
devices for camera, perception, and other always-on inference
tasks~\cite{qin2024mobilenetv4,ryali2023hiera}.
Real-time performance is critical for stable operation of these
workloads.

To meet real-time latency constraints, NPU compilers fuse consecutive
operators into monolithic
superlayers~\cite{chen2018tvm,rotem2018glow}.
This fusion can also create a few very large superlayers, which
concentrate computation into a short time window and cause large
current spikes.

Consequently, more aggressively fused, higher-efficiency designs may
lower average power yet create sharper current
spikes---an ``efficient'' model is not necessarily a ``stable'' one.
Through the power-delivery network (PDN), these bursts can induce
transient voltage droops~\cite{han2020pdnvoltdroop}; reduced
open-circuit voltage at low state of charge leaves less operating
margin~\cite{pattipati2014ocv}. In our voltage sweep, this regime is
associated with DVFS and shutdown-like behavior.
However, current mobile NPU compilation flows are effectively
voltage-unaware and have no
mechanism to detect or prevent this.
Meanwhile, extrinsic DVFS that suppresses instantaneous current can
only uniformly throttle the entire inference after the event,
incurring significant latency overhead.

\begin{figure*}[t]
  \centering
  \includegraphics[width=0.95\textwidth]{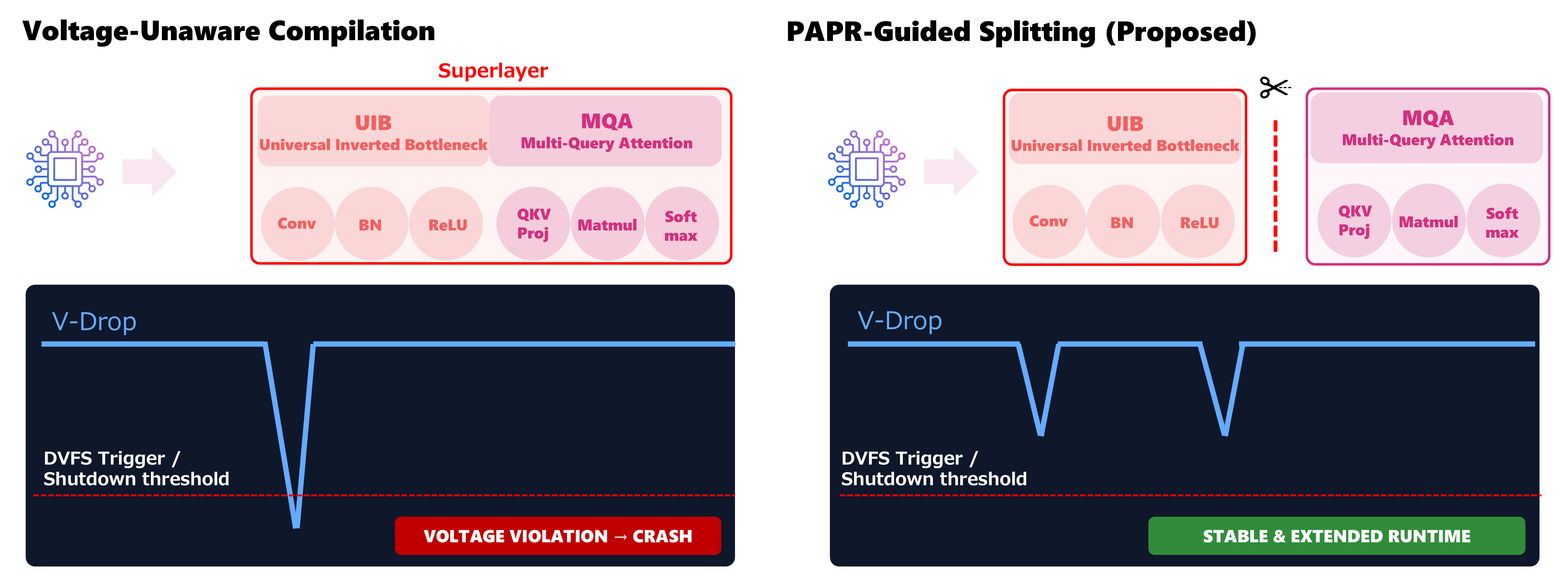}
  \caption{Voltage-unaware fusion and the proposed PAPR-based
  splitting.
  Left: The NPU compiler fuses operators (UIB + MQA,
  MobileNetV4) into a single superlayer to maximize throughput.
  The resulting current burst causes a deep voltage droop that crosses
  the DVFS\,/\,shutdown threshold.
  Right: The PAPR-based splitting method selectively splits the
  superlayer at identified boundaries, constraining each block's
  current peak within the safe voltage margin and extending stable
  operating time.}
  \label{fig:voltage_unaware_gap}
\end{figure*}

To address this problem, we propose a PAPR-based splitting method
(Fig.~\ref{fig:voltage_unaware_gap}):
before invoking the vendor NPU compiler, Q--DQ barrier layers
are inserted at selected fusion boundaries so that harmful merges do
not form overly large superlayers, thereby controlling peak current.

Our contributions are as follows:
\begin{enumerate}
  \item A measurement study on a commercial Snapdragon~8~Gen~3
    smartphone showing that mobile NPU compiler fusion can create
    peak-current hazards, and quantifying the relationship between
    peak current and DVFS activation voltage.
  \item A practical black-box mitigation based on PAPR-guided
    superlayer splitting, implemented as a pre-compilation graph
    rewrite that steers vendor compilation
    away from harmful merges under compiler constraints
    (Section~\ref{subsec:safety_layer}).
  \item Empirical design implications for mobile deployment, including
    operating regimes where splitting is worthwhile, comparison
    against vendor frequency capping, and results up to 38\,\% peak
    current reduction with $<$4\,\% latency increase
    (Section~\ref{sec:results}).
\end{enumerate}
 \section{Related Work}
\label{sec:related}

We organize prior work along two axes---compiler-side
(throughput-centric) and hardware-side (voltage-aware but generally
unaware of vendor compiler fusion decisions)---to motivate the gap
addressed by our measurement-guided graph rewrite.

\subsection{Compiler-side optimization}
NPU and DNN compilers pursue latency and memory efficiency through operator
fusion, tiling, and scheduling~\cite{chen2018tvm,rotem2018glow}.
In neural-architecture search (NAS),
NetAdapt~\cite{yang2018netadapt} and AMC~\cite{he2018amc} measure
per-layer latency;
FBNet~\cite{wu2019fbnet} builds operator-level latency
look-up tables;
Once-for-All~\cite{cai2020ofa} specializes sub-networks to
device-specific constraints.
On the energy side, Energy-Aware Pruning~\cite{yang2017energyaware}
targets inference energy as the objective, and HAQ~\cite{wang2019haq}
learns from hardware feedback on latency and energy.
MONAS~\cite{hsu2018monas} adds a scalar peak-power constraint to a
multi-objective NAS reward, but models peak power as a single number
rather than as a time-domain waveform.
In the compiler domain, Korch~\cite{zheng2024korch} employs operator
fission---decomposing operators into primitives before re-fusing
them---to discover faster kernel orchestrations;
notably, splitting is used solely to improve throughput, not to
mitigate voltage hazards.

Limitation.
These methods do not explicitly model time-domain PDN transients
induced by fusion or use measured current waveforms to veto selected
fusion boundaries.
Recent system-level work on DVFS governor tuning for LLM
inference~\cite{zhang2025dvfs} and energy-efficient early
exiting~\cite{zhang2025e4} likewise optimizes average energy without
modeling intra-inference voltage transients.

\subsection{Hardware-side mitigation}
At the physical layer, voltage droops arise from both resistive IR drop and
inductive $L\,\mathrm{d}I/\mathrm{d}t$ in the
PDN~\cite{han2020pdnvoltdroop,chang2017islped}.
Existing systems use DVFS and thermal throttling on embedded
processors~\cite{Peluso2019ElectronicsThermalDVFS}, while power capping
has been studied for scientific and multi-GPU
workloads~\cite{haidar2019cpe,krzywaniak2022iccs}.
These controls target power or temperature rather than
compiler-induced transient droop. Circuit-level techniques such as
Razor~\cite{ernst2003razor} and
active guardband management~\cite{Lefurgy2013ActiveGuardband} reduce
conservative margins.
Adaptive batching targets serving latency and
throughput~\cite{crankshaw2017clipper}, and heterogeneous DNN scheduling
targets responsiveness and energy~\cite{lin2024adaoper}. Conversely,
di/dt stressmarks deliberately generate periodic current pulses to
characterize worst-case PDN droop~\cite{Kim2011Stressmark}.
SparseDroop~\cite{raha2026sparsedroop} co-designs a dedicated hardware
stagger scheduler with structured weight pruning to reduce
$L\,\mathrm{d}I/\mathrm{d}t$ in custom DNN accelerators; however, it
requires dedicated hardware logic and model retraining, neither of
which is available on commodity mobile SoCs.

Limitation.
These mechanisms are extrinsic: they react to or prevent voltage
violations without understanding which compiler decisions caused them.
For example, DVFS uniformly lowers the clock for the entire inference,
incurring large latency overhead even when only one fused block is
responsible for the peak (see Section~\ref{sec:results},
Table~\ref{tab:burst_sustained}).

\subsection{Position of this work}
Our work targets the intersection of the two axes above on a commodity
mobile NPU.
We feed physical-layer measurements (current waveforms, PAPR) back into
a pre-compilation graph rewrite that selectively blocks only the
fusion boundaries responsible for voltage hazards.
In contrast to compiler-side methods---including Korch's operator
fission~\cite{zheng2024korch}---which split for throughput, our
method splits to reduce current density.
Our approach preserves weights and is deployable on commodity devices.
The resulting PAPR-based splitting method is orthogonal to,
and composable with, both compiler optimizations and extrinsic
throttling.

 \section{Method}\label{sec:method}

We first define the PDN transient model and show how voltage-unaware
compiler fusion creates voltage hazards.
We then describe the details of the PAPR-based splitting algorithm.

\subsection{Voltage hazards from voltage-unaware fusion}\label{subsec:voltage_unaware_gap}

\begin{figure}[t]
  \centering
  \includegraphics[width=0.5\linewidth]{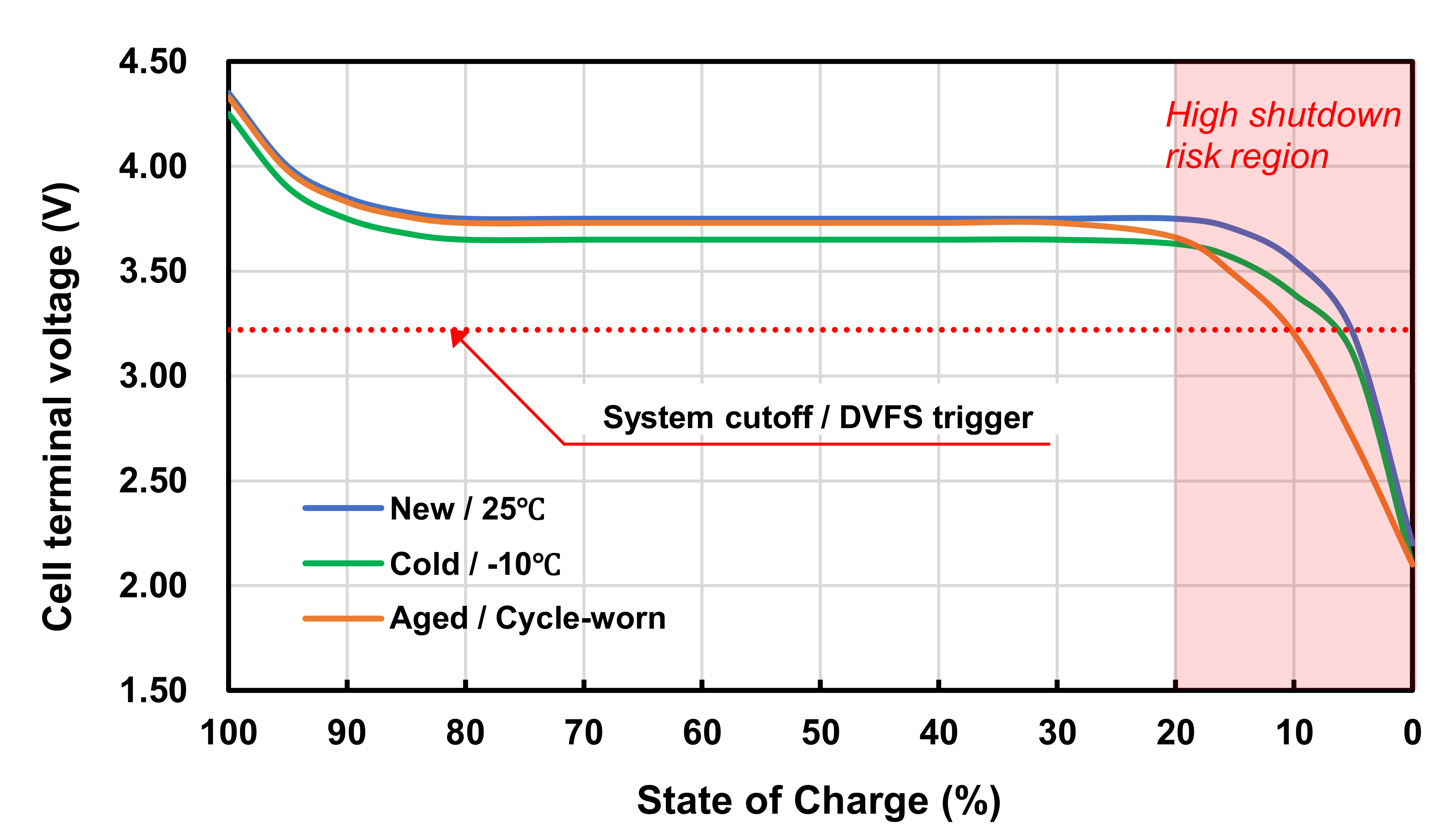}
  \caption{Li-ion terminal voltage vs.\ SOC (schematic, based
  on~\cite{pattipati2014ocv,xiong2018soc}).
  At low SOC, $V_{\mathrm{src}}$ is small and peak-induced
  $\Delta V$ readily pushes $V_{\mathrm{term}}$ below the DVFS
  or shutdown threshold.}
  \label{fig:liion_soc}
\end{figure}

\paragraph{PDN transient model.}
We model the device's power-delivery network as a series resistance
$R$ and inductance $L$.
Given load current $I(t)$, the instantaneous voltage drop on the supply rail
is
\begin{equation}
  \Delta V(t) \;\approx\; R\,I(t) \;+\; L\,\frac{\mathrm{d}I(t)}{\mathrm{d}t}\,,
  \label{eq:pdn_drop}
\end{equation}
and the SoC terminal voltage is
\begin{equation}
  V_{\mathrm{term}}(t) \;=\; V_{\mathrm{src}}(t) \;-\; \Delta V(t)\,,
  \label{eq:vterm}
\end{equation}
where $V_{\mathrm{src}}(t)$ is the battery open-circuit voltage, which
drops steeply at low state of
charge~\cite{pattipati2014ocv,xiong2018soc}
(Fig.~\ref{fig:liion_soc}).
When $V_{\mathrm{term}}$ falls below a hardware threshold, low-voltage
protection can engage even while the displayed SoC remains above the
software-protection threshold.

\paragraph{How compiler fusion creates voltage hazards.}
NPU compilers fuse consecutive operators into superlayers,
eliminating intermediate buffers and maximizing
throughput~\cite{chen2018tvm,rotem2018glow}.
This is optimal in the compiler's logical domain.
However, in the physical domain, each superlayer concentrates
compute and memory traffic into a short window, producing
(i)~high $|I(t)|$, which increases the instantaneous rail drop, and
(ii)~large localized power bursts relative to the inference average.
Wide activations and channel expansions further amplify simultaneous
memory traffic and MAC bursts~\cite{horowitz2014energy}.

The target compiler has no model of these physical consequences; it is
voltage-unaware.
The result is that fusion decisions locally optimal for throughput
become globally harmful for voltage stability.

\subsection{PAPR-based splitting}\label{subsec:safety_layer}

To counteract the voltage hazards created by voltage-unaware fusion,
we apply a graph-rewrite pass before invoking the vendor NPU compiler.
The deployed transformation is lightweight and performs no online
search: it inserts up to $K$ barriers at PAPR-ranked boundaries to block
the largest voltage-hazard superlayers. Constructing the feasible-edge
set and measuring the PAPR profile are one-time offline characterization
steps for each model--resolution--compiler configuration.

\paragraph{Peak metric: PAPR.}
We quantify peaks using the peak-to-average power ratio (PAPR).
Let $I(t)$ be the instantaneous current at time~$t$ and $\bar{I}$
the mean current over the full inference interval; then
\begin{equation}
  \mathrm{PAPR}(t) \;=\; \frac{I(t)}{\bar{I}}
  \label{eq:papr_definition}
\end{equation}
is a dimensionless measure of instantaneous stress relative to the
inference-average current on the measured device.
Because the supply-rail voltage is approximately constant during a
single inference, PAPR in current is proportional to conventional
power PAPR, making it a convenient proxy for instantaneous power
stress. In this work, we use PAPR as a practical peak-current
indicator for identifying superlayers that are likely to create the
largest droop on the measured device.

\paragraph{Compile-feasible edge set.}
Let $E$ contain every edge in the exported graph. For each
model--resolution--compiler configuration, we tentatively insert a Q--DQ
pair at every $b\in E$, export the rewritten graph, invoke the vendor
compiler, and test the artifact on the target NPU. We define
\begin{equation}
  F = \{\,b\in E \mid \textsc{CompileFeasible}(b)\,\},
  \label{eq:feasible_set}
\end{equation}
where \textsc{CompileFeasible} requires successful graph export, vendor
compilation, and execution. Exhaustive probing is necessary because
some edges that appear structurally insertable fail vendor compilation
when a Q--DQ pair is placed inside the corresponding fused region. The
set $F$ is computed once offline for each configuration and used as an
internal compatibility-screening pool. The deployable output is the
jointly compiled boundary set rather than the pool of individually
feasible edges. For exploratory characterization, we report the final
sets $B_{\mathrm{eval}}$ in Appendix Table~\ref{tab:s_boundary_set}; at
deployment, Algorithm~\ref{alg:minimal_split_alg} returns $B^*$ after
applying the PAPR gate.

\paragraph{Greedy boundary selection.}
Given $F$ and the PAPR profile, Algorithm~\ref{alg:minimal_split_alg}
first identifies compile-feasible boundaries adjacent to PAPR hot spots,
defined by $\mathrm{PAPR}\ge\gamma$. It then ranks these candidates by
local peak current and greedily adds up to $K$ boundaries. For an edge
$b=(i,j)$, its boundary interval is $T_b=[t_s^i,t_e^j]$ and its score is
$s(b)=\max_{t\in T_b}I(t)$, using the synchronized execution intervals of
its producer and consumer.

\begin{algorithm}[H]
\small
\caption{PAPR-Guided Greedy Split}
\label{alg:minimal_split_alg}
\begin{algorithmic}[1]
\Require Current waveform $I(t)$; layer intervals
  $\{[t_s^{i},\, t_e^{i}]\}_{i=1}^{N}$;
  compile-feasible edge set $F$; threshold $\gamma$; budget $K$
\Ensure Split boundaries $B^*$
\State $S \gets \{\,t \mid I(t)/\bar{I} \ge \gamma\,\}$
  \Comment{PAPR hot points}
\If{$S = \emptyset$} \Return $\emptyset$ \EndIf
\State $C \gets \{b\in F \mid T_b \cap S \ne \emptyset\}$
  \Comment{feasible hot-spot boundaries}
\State Sort $C$ by $s(b)$, descending
\State $B^* \gets \emptyset$
\For{$b$ in sorted $C$}
  \If{$|B^*|<K$ \textbf{and} $\textsc{JointCompileFeasible}(B^*\cup\{b\})$}
    \State $B^* \gets B^*\cup\{b\}$
  \EndIf
\EndFor
\State \Return $B^*$
\end{algorithmic}
\end{algorithm}

\textsc{JointCompileFeasible} requires successful graph export, vendor
compilation, and execution with all barriers in the candidate set. A
candidate that fails this combined check is skipped.

At each selected boundary in $B^*$, we insert a Quantize--Dequantize
(Q--DQ) operator pair as a barrier that the compiler cannot fold
away. In principle, a zero-cost identity operator would be more
desirable. In practice, however, the target NPU compiler (Qualcomm
HTP) merges Identity and Reshape nodes, defeating the intended split;
we therefore use Q--DQ as a practical substitute. The rewritten graph
is then passed to the vendor compiler, which sees the separated
subgraphs and no longer merges them into a single monolithic
superlayer.
The measured accuracy impact is reported in
Table~\ref{tab:k_ablation} (worst case $-0.31$\,pp Top-1).

The PAPR threshold $\gamma$ is an operational parameter that defines
which superlayers are treated as voltage hazards. Its value should be
chosen according to the device's DVFS-onset characteristics and the
trade-off between peak reduction and latency overhead. In
Section~\ref{sec:results}, we derive an empirical
operating threshold from measurements across models and resolutions.
The parameter $K$ sets the
maximum number of inserted barriers and therefore directly limits the
latency overhead of the method.
 \section{Experimental Setup}
\label{sec:experiments}

This section describes the measurement environment, the voltage-sweep
protocol used to expose droop-induced protection behavior, the models to
which the proposed splitting method is applied, and the evaluation
metrics used throughout the paper.

\subsection{Measurement setup}
\label{subsec:motivating_experiment}

\begin{figure*}[t]
    \centering
    \includegraphics[width=0.75\linewidth]{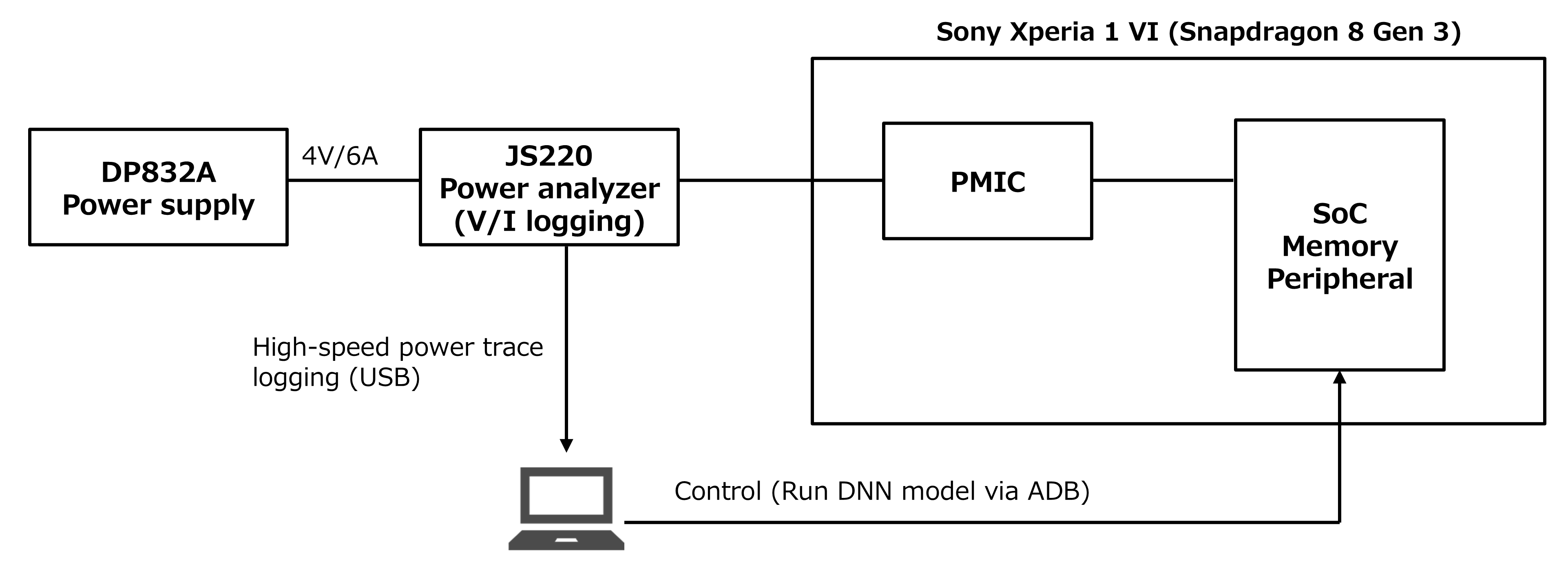}
    \caption{Power measurement setup.
    The battery terminals are driven by an external supply (DP832A),
    and instantaneous power at the PMIC input rail is acquired with
    a power analyzer (JS220).
    A host PC synchronously collects DNN inference logs and power
    traces via Android Debug Bridge (ADB).}
    \label{fig:measurement_setup}
\end{figure*}

To confirm that compiler-fused superlayers create measurable voltage
hazards, we synchronously measured current, voltage, and latency during
DNN inference on a commercial smartphone (Sony Xperia~1~VI,
Snapdragon~8~Gen~3, equipped with a Qualcomm Hexagon Tensor Processor
(HTP) NPU; Fig.~\ref{fig:measurement_setup}).

To isolate the NPU contribution, we suppressed background processing,
fixed CPU and DRAM frequencies to their lowest values, and ensured that
only the NPU was under high load. Subtracting idle power from active
power yielded the NPU load power.

\subsection{Voltage-sweep protocol and protection modes}
\label{subsec:shutdown_protection}

To characterize DVFS protection and shutdown-like behavior, we ran
continuous inference while lowering the programmable DC supply at the
PMIC input rail in $\sim$10\,mV steps, recording the
DVFS incidence (fraction of throttled inference kernels) and latency
at each voltage point.
For the DVFS-onset analysis, each peak-current bin contains five
measurements. In the detailed MobileNetV4@768 voltage sweep, latency at
each plotted point is averaged over 100 inferences.
Waits were inserted between frames to equalize average power across
models while keeping each DNN's instantaneous peak current distinct.
This sweep revealed how the device's protection mechanisms engage.
All measurements were performed at 25\,$^\circ$C with sufficient
cool-down between voltage points.

In this evaluation, we distinguish two protection mechanisms:

\subsubsection{Software protection}
During normal operation, the OS estimates a safety margin from the
indicated state-of-charge (SoC) and uniformly enforces DVFS in
software when SoC approaches about 5\%.
In this mode, throttling fires irrespective of the model's peak
power once SoC nears the threshold, causing a sharp latency
increase (up to 72\% in our measurements).
Thus, software protection is an SoC-based, model-agnostic
throttling policy rather than a response to model-specific power
profiles.

\subsubsection{Hardware protection}
Beyond software control, if the effective PMIC input voltage
$V_{\mathrm{term}}(t)$ momentarily drops below a hardware
threshold due to PDN-induced droop
(Section~\ref{subsec:voltage_unaware_gap}), the device engages
DVFS or near-halt behavior as a self-defense mechanism.
This response is independent of the displayed SoC and is driven
directly by instantaneous current spikes and the resulting voltage
drop.
By replacing the internal battery with a programmable supply, we
could lower the supply voltage even while the device judged
sufficient capacity to remain (i.e., before software protection
would trigger), allowing us to isolate the impact of voltage droop
during DNN inference.

\subsection{Models}
\label{subsec:peak_mitigation}
We apply Algorithm~\ref{alg:minimal_split_alg} to
MobileNetV4-hybrid-large~\cite{qin2024mobilenetv4} and
Hiera-Tiny~\cite{ryali2023hiera}. These two backbones were chosen to
cover representative mobile vision architectures with different
operator structures and fusion behavior on the target NPU.

\subsection{Evaluation metrics}
\label{subsec:eval_metrics}
For each model and resolution, we report the following before and
after splitting:
{\renewcommand{\baselinestretch}{0.93}\selectfont
\begin{itemize}
  \item Latency: per-inference processing time [ms].
  \item Current (MAX): maximum instantaneous current within the
        recorded inference interval at the PMIC input rail [A].
  \item Avg Power: mean power during inference [W].
  \item Energy / inference: Latency $\times$ Avg Power [mJ].
\end{itemize}
}
These metrics capture both the intended benefit (peak reduction) and
the cost (latency and energy overhead) of breaking fusion.
 \section{Results and Discussion}
\label{sec:results}

This section first characterizes the association between instantaneous
peak current and DVFS onset, then evaluates the proposed
splitting method across models and input resolutions, and finally
discusses the practical implications and limitations of the approach.

\subsection{Peak current and hardware protection threshold}

\begin{figure}[t]
  \centering
  \includegraphics[width=0.5\linewidth]{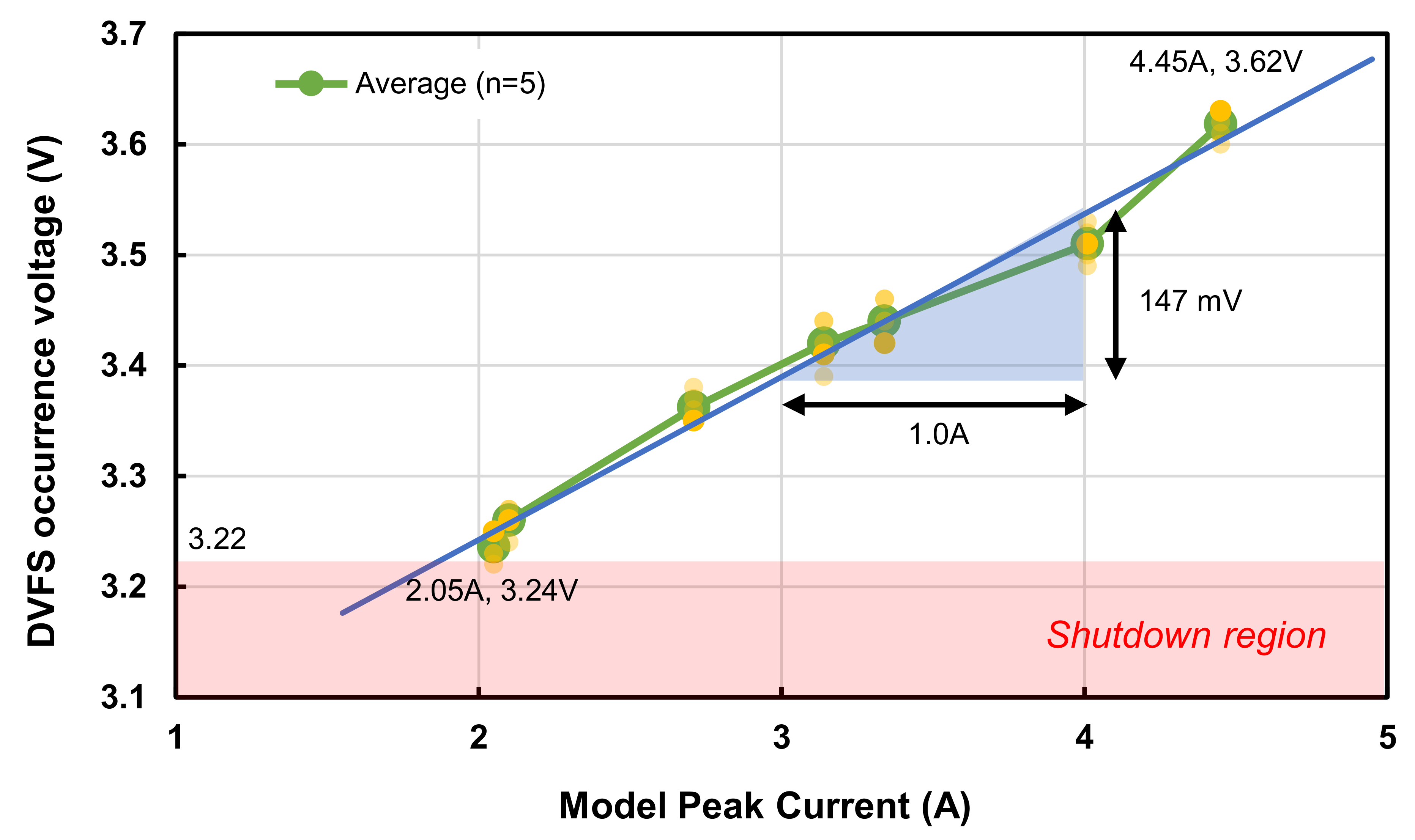}
  \caption{DVFS-onset voltage vs.\ peak current.
  Dots: all measurements per peak-current bin ($n{=}5$).
  Thick line: bin-wise mean.
  Dashed line: exploratory linear fit (extrapolated), slope 0.147\,V/A
  ($\approx$147\,mV per +1\,A).
  The orange band marks the shutdown-dominated region at
  $V_{\mathrm{LVC}} \approx 3.22$\,V.
  Peak reduction by the PAPR-based splitting method
  (Section~\ref{subsec:safety_layer}) is estimated to lower the
  DVFS-onset voltage toward the shutdown threshold.}
  \label{fig:voltage_dvfs_ratio}
\end{figure}

Fig.~\ref{fig:voltage_dvfs_ratio} shows the relationship between peak
current and DVFS-onset voltage for each model. Models with higher peak
current exhibit DVFS onset at higher supply voltages; an exploratory
linear fit has a slope of 0.147\,V/A. Inserting waits between frames
reduces differences in average load across models, but it does not
eliminate model- and kernel-dependent confounders. We therefore
interpret the fit as an association rather than an isolated causal
estimate of peak current.

The trend is consistent with peak-current-induced PDN drop advancing
hardware-protection DVFS onset.

\subsection{Qualitative example of PAPR-guided splitting}

\begin{figure*}[t]
  \centering
  \includegraphics[width=1.0\textwidth]{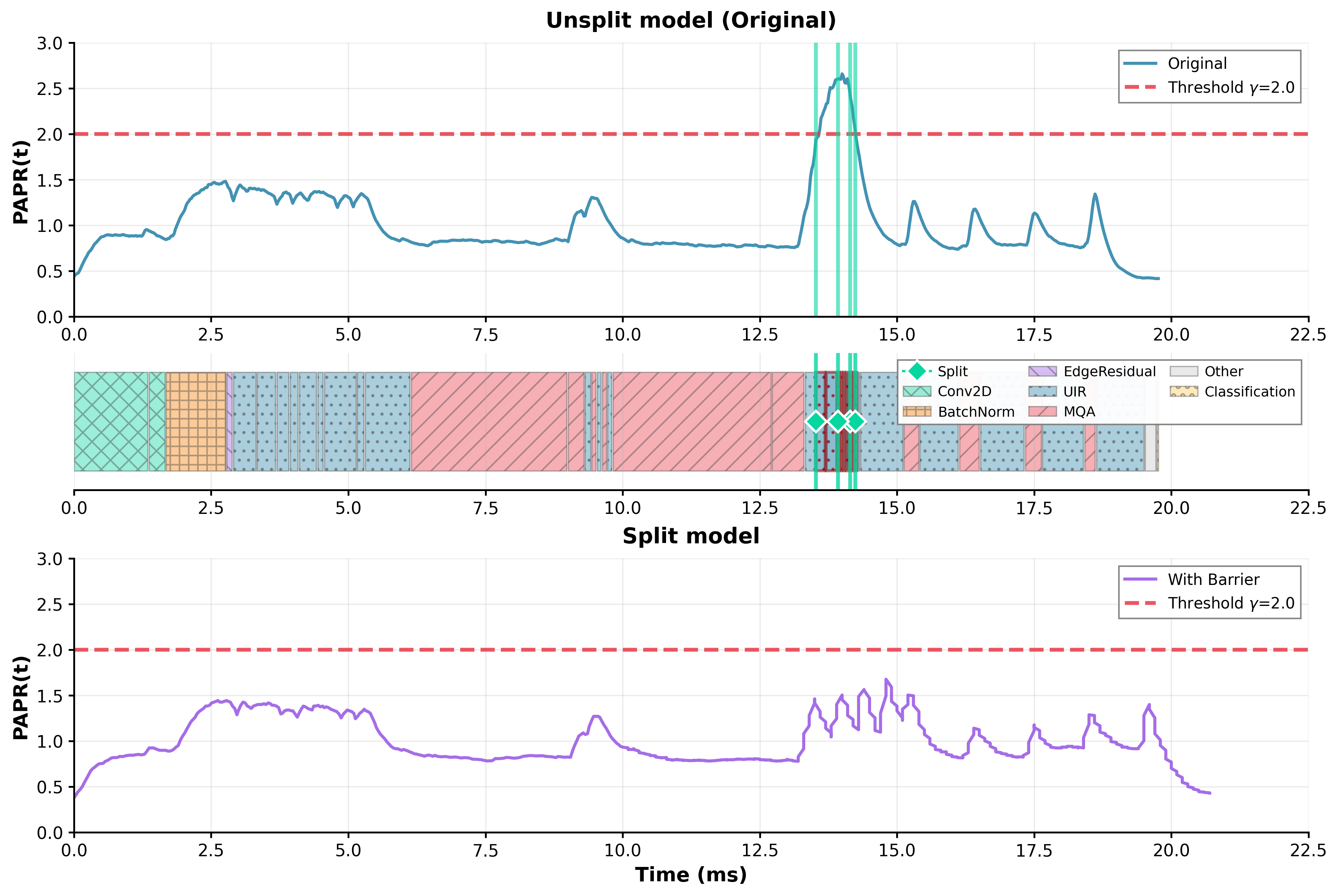}
  \caption{Correspondence between PAPR (upper) and layer execution
  intervals (lower) in MobileNetV4 (768$\times$768 input).
  The upper panel shows
  $\mathrm{PAPR}(t) = I(t)/\bar{I}$ relative to the average
  current $\bar{I}$ over the entire inference interval; the red
  dashed line marks the threshold $\gamma = 2.0$.
  The middle panel color-codes per-layer execution intervals.
  Light blue vertical lines indicate hot intervals exceeding
  $\gamma$, and green diamonds are split boundaries selected by
  the algorithm ($K{=}4$ splits).
  The large Conv2D region (red diagonal lines) with consecutive
  UIR blocks is fused as a single superlayer, generating a PAPR
  peak ($\approx$2.6).
  The lower panel shows the split model after applying our method,
  where the PAPR peak is reduced below the threshold $\gamma$.}
  \label{fig:mnv4_layer_decomposition}
\end{figure*}

On the Sony Xperia~1~VI, the sequence of Universal Inverted Residual
(UIR) blocks in MobileNetV4 is fused by the NPU compiler into a single
large superlayer, producing a pronounced peak-power burst
(Fig.~\ref{fig:mnv4_layer_decomposition}). This superlayer arises
because multiple UIR blocks share the same spatial resolution, kernel
shape, and operator pattern (Conv $\rightarrow$ BN $\rightarrow$ Act,
etc.), prompting the NPU optimizer to schedule them as one large fused
kernel.

\subsection{Main quantitative results}

We first performed an exploratory characterization to determine when
splitting is worthwhile, without applying a PAPR-based deployment gate.
For each model--resolution pair, we tested Q--DQ insertion at every graph
edge $b\in E$ and formed the compile-feasible subset $F$ using
Eq.~\eqref{eq:feasible_set}. We then disabled the deployment threshold,
ranked the screened edges by local peak current, and constructed a jointly
compilable evaluation set $B_{\mathrm{eval}}\subseteq F$ subject to
$|B_{\mathrm{eval}}|\le K=4$, using the same greedy skip-on-failure rule
as Algorithm~\ref{alg:minimal_split_alg}. This forced-split
characterization includes low-PAPR cases and exposes when insertion
overhead outweighs peak-current reduction. The resulting sets contain
four barriers for MobileNetV4 and two for Hiera-Tiny; their locations and
detailed operator listings are reported in the appendix.

\begin{table*}[t]
\centering
\footnotesize
\caption{Comparison of latency, peak current, average power, PAPR,
and energy across architectures.
For each model and resolution, the third row reports ratios
(Forced/Original). The forced-split rows are exploratory results obtained
without applying the deployment gate.
Resolutions 224--1024 denote input size.
\textbf{Bold} marks the selected operating point that achieves over
35\% peak reduction with less than 4\% latency overhead.}\label{tab:main_results}
\resizebox{\textwidth}{!}{\begin{tabular}{@{}llcccccc@{}}
\hline
Model & Resolution & Variant & Latency [ms] & Current (MAX) [A] & Avg Power [W] & PAPR$_{\max}$ & Energy / inference [mJ] \\
\hline
\multirow{12}{*}{MobileNetV4} & 224$\times$224 & Original & 2.04 & 1.99 & 4.69 & 1.57 & 9.57 \\
                              &                & Forced split & 2.56 & 1.54 & 4.44 & 1.28 & 11.36 \\
                              &                & Ratio (Forced/Orig.) & 1.25$\times$ & 0.77$\times$ & 0.95$\times$ & & 1.19$\times$ \\
\cline{2-8}
                              & 512$\times$512 & Original & 6.18 & 3.06 & 4.65 & 2.43 & 28.73 \\
                              &                & Forced split & 7.07 & 1.85 & 4.60 & 1.49 & 32.52 \\
                              &                & Ratio (Forced/Orig.) & 1.14$\times$ & 0.60$\times$ & 0.99$\times$ & & 1.13$\times$ \\
\cline{2-8}
                              & 768$\times$768 & Original & 19.95 & 3.12 & 4.44 & 2.60 & 88.58 \\
                              &                & Forced split & 20.70 & 1.94 & 4.39 & 1.64 & 90.87 \\
                              &                & \textbf{Ratio (Forced/Orig.)} & \textbf{1.04$\times$} & \textbf{0.62}$\times$ & \textbf{0.99}$\times$ & & \textbf{1.03}$\times$ \\
\cline{2-8}
                              & 1024$\times$1024 & Original & 47.63 & 3.04 & 4.32 & 2.60 & 205.76 \\
                              &                  & Forced split & 48.42 & 2.51 & 4.38 & 2.12 & 212.08 \\
                              &                  & Ratio (Forced/Orig.) & 1.02$\times$ & 0.83$\times$ & 1.01$\times$ & & 1.03$\times$ \\
\hline
\multirow{6}{*}{Hiera-Tiny}  & 224$\times$224 & Original & 6.29 & 1.74 & 4.22 & 1.53 & 26.54 \\
                              &                & Forced split & 8.54 & 1.36 & 3.10 & 1.62 & 26.47 \\
                              &                & Ratio (Forced/Orig.) & 1.36$\times$ & 0.78$\times$ & 0.73$\times$ & & 1.00$\times$ \\
\cline{2-8}
                              & 384$\times$384 & Original & 26.33 & 2.07 & 4.60 & 1.66 & 121.12 \\
                              &                & Forced split & 32.90 & 2.06 & 3.98 & 1.91 & 130.94 \\
                              &                & Ratio (Forced/Orig.) & 1.25$\times$ & 0.99$\times$ & 0.86$\times$ & & 1.08$\times$ \\
\hline
\end{tabular}
}
\end{table*}

\begin{table}[t]
\centering
\footnotesize
\caption{Effect of Q--DQ layer insertion count $K$ on accuracy
(MobileNetV4-hybrid-large, timm pretrained, ImageNet-1k,
448$\times$448 input). For $K>0$, barriers are accumulated in descending
local-peak rank using the same skip-on-failure rule as Algorithm~1.}
\label{tab:k_ablation}
\begin{tabular}{c|cc|cc}
\toprule
\multirow{2}{*}{$K$} & \multicolumn{2}{c|}{Accuracy [\%]} & \multicolumn{2}{c}{$\Delta$ vs. $K{=}0$ [\%]} \\
\cline{2-5}
 & Top-1 & Top-5 & Top-1 & Top-5 \\
\midrule
0 (none) & 84.25 & 96.94 & \multicolumn{2}{c}{---} \\
1         & 83.94 & 96.82 & \textbf{-0.31} & -0.11 \\
2         & 84.03 & 96.94 & -0.22 & {\small +0.00} \\
3         & 84.06 & 96.93 & -0.20 & -0.01 \\
4         & 84.06 & 96.91 & -0.19 & -0.03 \\
\bottomrule
\end{tabular}
\end{table}

Focusing on the $\mathrm{PAPR}_{\max}$ column in
Table~\ref{tab:main_results}, the six configurations separate into two
empirical regimes. For the three configurations with
$\mathrm{PAPR}_{\max} \ge 2.0$
(MNV4@512: 2.43, @768: 2.60, @1024: 2.60), forced splitting reduces peak
current by 17--40\% with only 2--14\% latency increase. MNV4@768 is a
particularly favorable operating point, with 38\% peak reduction at
4\% latency cost.

In contrast, for the three configurations with
$\mathrm{PAPR}_{\max} < 2.0$
(MNV4@224: 1.57, Hiera@224: 1.53, Hiera@384: 1.66), forced-split peak reduction is
limited to 0--23\% while latency increases by 25--36\%---the
barrier-insertion overhead exceeds the benefit of peak reduction
because the original peak is already low.

This exploratory dichotomy suggests an empirical operating threshold near
$\mathrm{PAPR}_{\max} \approx 2.0$ on the tested platform. We therefore
use $\gamma=2.0$ as a deployment gate: configurations below the threshold,
including both Hiera-Tiny cases, are left unsplit, whereas candidate
boundaries are considered only when the measured profile crosses the
threshold. The Hiera-Tiny results are therefore low-PAPR cases showing
why unconditional splitting is undesirable, not configurations for which
the final policy recommends splitting. Accuracy is also largely
preserved: the largest Top-1
change with Q--DQ layer insertion is $-0.31$\,pp
(Table~\ref{tab:k_ablation}). That said, as noted earlier, an insertion
layer that is computationally equivalent or does not affect accuracy
would be ideal for layer splitting.

\subsection{Sensitivity to boundary scoring}
\label{subsec:sensitivity}

Under the conditions $\gamma{=}2.0$ and $K{=}4$ identified as effective
in the trade-off analysis above, we performed splitting using three patterns:
PAPR-threshold selection, equal-interval splitting, and random
splitting. All three strategies select from the same compile-feasible
candidate pool under the same $K=4$ budget: equal-interval splitting
spaces its choices uniformly in execution order, while random splitting
samples from the pool. For every strategy, candidates that failed the
same combined compilation check were skipped. The experiment was
conducted on MobileNetV4@768, comparing peak current and latency
(Table~\ref{tab:baseline_comparison}).
PAPR-threshold selection reduced peak current to 1.94\,A, whereas
equal-interval splitting achieved only 2.51\,A and random splitting
averaged 2.65\,A, both requiring comparable or higher latency. This
demonstrates that PAPR-based boundary selection is essential for peak
reduction.

\begin{table}[t]
\centering
\footnotesize
\caption{Comparison of boundary scoring strategies
(MobileNetV4@768, budget $K{=}4$).
PAPR-guided selection reduces peak current 23--27\% more than
na\"ive methods.
Random Split values are averaged over 3 seeds.}
\label{tab:baseline_comparison}
\begin{tabular}{l@{\hspace{2mm}}c@{\hspace{2mm}}c}
\toprule
Method & Peak Current (A) & Latency (ms) \\
\midrule
Proposed (PAPR) & 1.94 & 20.70 \\
Equal Interval & 2.51 & 21.17 \\
Random ($s{=}0,1,2$) & 2.65 avg (2.50--2.74) & 20.88 (20.70--21.10) \\
\bottomrule
\end{tabular}
\end{table}

\subsection{Implications for voltage margin and control}

\paragraph{Voltage-margin improvement.}
As discussed in Section~\ref{subsec:shutdown_protection}, we estimate
the voltage-margin change associated with peak-current reduction. From
the exploratory fit in Fig.~\ref{fig:voltage_dvfs_ratio}, the DVFS
activation voltage changes by approximately 147\,mV per 1\,A of peak
current. The 1.18\,A reduction for MobileNetV4@768
(3.12\,A $\rightarrow$ 1.94\,A) therefore yields an improvement in the
DVFS activation voltage of approximately 173\,mV:
\begin{equation}
\Delta V_{\mathrm{DVFS}} = 1.18\,\mathrm{A} \times 147\,\mathrm{mV/A} \approx 173\,\mathrm{mV}
\end{equation}

\begin{figure}[t]
  \centering
  \includegraphics[width=0.5\linewidth]{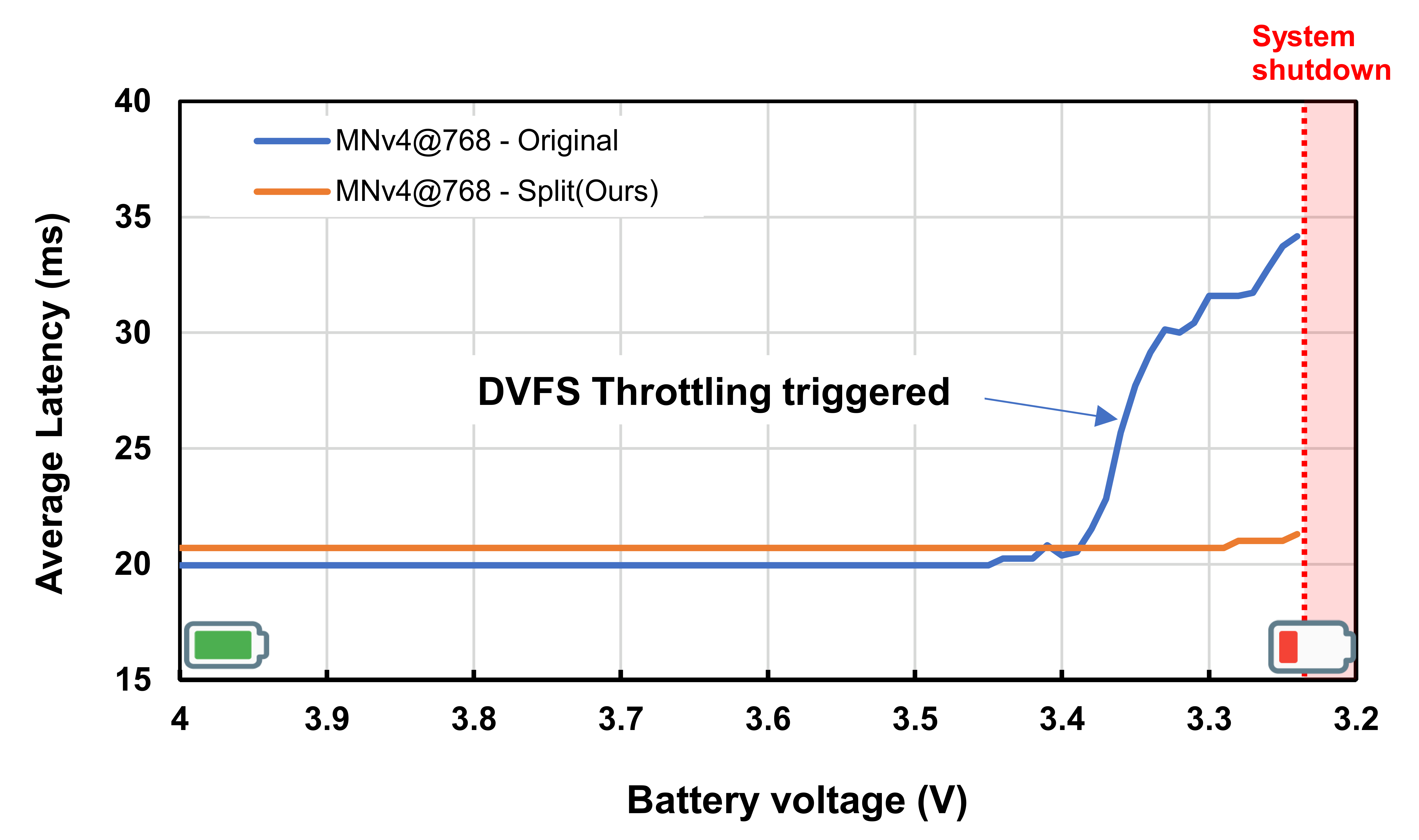}
  \caption{Applied voltage vs.\ average latency
  (MobileNetV4@768).
  Inference time is the average over every 100 inferences.
  The Original version shows sharp latency increases from DVFS
  throttling around 3.45\,V, while the Split version maintains
  stable inference time down to near the shutdown threshold
  ($\approx$3.28\,V).}
  \label{fig:voltage_latency_split}
\end{figure}

Fig.~\ref{fig:voltage_latency_split} shows the corresponding within-model
comparison. We deployed MobileNetV4@768 before and after applying the
algorithm on a real device and measured average latency at each applied
voltage. The Original version begins to exhibit latency increases due
to DVFS throttling around 3.45\,V, whereas the Split version maintains
stable latency down to near the system shutdown threshold
($\approx$3.28\,V). This comparison is consistent with peak-current
reduction shifting hardware-protection DVFS activation to a lower
voltage.

\paragraph{Cross-device interpretation.}
The slope in Fig.~\ref{fig:voltage_dvfs_ratio} is a descriptive estimate
of the effective current-to-voltage sensitivity of the measured
power-delivery path. We do not infer its magnitude on other devices;
cross-device use requires a new fit under matched acquisition settings.

As shown in Fig.~\ref{fig:voltage_dvfs_ratio}, in our measurements the
inference task became unsustainable (OS shutdown) once the terminal/PMIC
input voltage reached approximately 3.22\,V. Reducing the peak current
shifts the DVFS activation voltage downward by $\Delta V$, which
effectively widens the usable operating-voltage range under the same
load profile.

\subsection{Comparison with frequency capping}

\begin{table*}[!t]
\centering
\footnotesize
\captionsetup{font=footnotesize}
\caption{Comparison with vendor frequency capping: PAPR-based splitting vs.\
frequency capping (MobileNetV4@768).
Sustained mode is a vendor preset of frequency capping.
PAPR follows Eq.~\eqref{eq:papr_definition}:
$\mathrm{PAPR}(t) = I(t)/\bar{I}$.}\label{tab:burst_sustained}
\begin{tabular*}{\textwidth}{@{\extracolsep{\fill}}llcccccccc@{}}
\hline
Method & NPU mode & Lat. [ms] & $\Delta$Lat. [\%] & Peak [A] & $\Delta$Peak [\%] & Power [W] & Energy [mJ] & $\Delta$Energy [\%] \\
\hline
Baseline & Burst & 19.95 & 0 & 3.12 & 0 & 4.44 & 88.58 & 0 \\
Baseline & Sustained & 23.73 & +18.9 & 2.29 & $-26.6$ & 3.35 & 79.50 & $-10.3$ \\
\textbf{Proposed} & \textbf{Burst} & \textbf{20.70} & \textbf{+3.8} & \textbf{1.94} & \textbf{$-$37.8} & \textbf{4.39} & \textbf{90.87} & \textbf{+2.6} \\
Proposed & Sustained & 25.04 & +25.5 & 1.49 & $-52.2$ & 3.20 & 80.13 & $-9.5$ \\
\hline
\end{tabular*}
\end{table*}

The proposed method reshapes the current waveform at the compiler level
and is therefore orthogonal to external frequency-capping strategies
such as DVFS (Section~\ref{sec:related}). To compare against a practical
external-control baseline, we use the vendor-provided frequency cap:
Qualcomm's QNN HTP backend exposes a ``perf\_profile'' knob with
several presets (e.g., ``burst'', ``sustained\_high\_performance''),
which directly limit the available clock/voltage operating points. We
treat sustained mode as a representative vendor preset of frequency
capping and compare it against PAPR-based splitting on
MobileNetV4@768 (Table~\ref{tab:burst_sustained}).

Frequency capping (sustained mode) lowers the absolute peak but at a
large latency cost ($\times$1.19 vs.\ baseline burst). In contrast,
PAPR-based splitting achieves a larger peak reduction
($-37.8$\% vs.\ $-26.6$\%) with only $+3.8$\% latency in burst mode.
Critically, frequency capping reduces the absolute current peak but
does not reshape the waveform as selectively as the proposed method:
its $\mathrm{PAPR}_{\max}$ changes only modestly
(2.60$\rightarrow$2.12), whereas the proposed method reduces it more
substantially (2.60$\rightarrow$1.64) by breaking only the
highest-hazard superlayers. The two approaches are complementary:
combining them (Proposed~+~Sustained) yields a 52.2\% peak reduction.

\subsection{Limitations}

This paper is a measurement study on one SoC (Snapdragon~8~Gen~3 NPU)
and two representative vision backbones. The specific fusion patterns of
the NPU compiler examined in this paper are expected to vary across
compilers and devices. The method is most effective when the compiler
creates large, aggressively fused superlayers; when fusion is minimal,
the effect of the safety layer is limited and conventional
energy/latency optimizations remain the primary lever. Results also vary
with input resolution, as different resolutions change NPU kernel
selections and memory-access patterns.
The current workflow requires offline PAPR profiling and compilation
screening for each model--resolution--compiler configuration.

The voltage-onset regression is descriptive: we report the
sample count and fitted slope, but no confidence interval or hypothesis
test. Accordingly, the 147\,mV/A slope and the derived 173\,mV margin are
exploratory estimates rather than calibrated transfer laws.

Furthermore, on CPU/GPU delegates that do not perform aggressive layer
fusion for execution-speed reduction, superlayers themselves are
unlikely to form, and the effect of the proposed method is expected to
be limited. Accordingly, we do not claim that the measured DVFS slope
or the best splitting threshold transfers unchanged across devices;
rather, we view these results as evidence that
compiler-fusion-induced power transients are a practically measurable
deployment issue that merits per-platform characterization.
 \section{Conclusion}
\label{sec:conclusion}

This paper reported a measurement study of compiler-fusion-induced
voltage droop during mobile NPU inference.
Aggressive operator fusion maximizes throughput in the logical domain
but can create high-amplitude current bursts that threaten voltage
stability in the physical domain.
During a synchronized current measurement and voltage sweep on a
Snapdragon~8~Gen~3 smartphone, these bursts were associated with
earlier DVFS onset and shutdown-like behavior near the lower operating
limit.

To counteract these voltage hazards, we proposed a PAPR-based
splitting method---a measurement-guided, weight-preserving workflow
whose deployed rewrite selectively splits the highest-hazard
superlayers using PAPR-guided boundary selection.
On MobileNetV4 (768$\times$768), peak current was reduced by
37.8\,\% ($-$1.18\,A) with only 3.76\,\% latency overhead,
yielding an estimated 173\,mV improvement in DVFS voltage margin on
the tested device.
PAPR-guided selection outperformed random and equal-interval
baselines by 23--27\,\% in peak reduction, confirming that
targeted boundary identification is essential.

These results highlight a practical deployment issue for always-on
mobile inference pipelines---including camera and perception
workloads---where DVFS-induced latency spikes cause frame drops and
degrade user experience in the low-battery regime.
The proposed safety layer complements both model compression/NAS and
extrinsic throttling, and is best suited to reshaping current
waveforms without retraining, modifying the compiler, or
compromising hardware protection.
More broadly, we expect the measurements in this study to be useful
as a reference point for future work on mobile NPU compilers, power
integrity, and low-voltage inference robustness.

In future work, we will extend the approach to additional SoCs,
compilers, and model families, and investigate static peak-power
prediction from operator-graph features to eliminate the need for
per-model measurement.
 \appendix
\clearpage
\renewcommand{\thetable}{A\arabic{table}}
\setcounter{table}{0}

\section{Supplementary Material}
\label{sec:supplementary_material}

This appendix reports the exploratory boundary sets and their
operator-level locations.

\begin{table*}[!ht]
    \centering
    \footnotesize
    \caption{Boundary sets selected from the compile-feasible set $F$
    for exploratory forced-split characterization. Indices refer to the
    rewritten graph.}
    \label{tab:s_boundary_set}
    \begin{tabular}{lccc}
        \toprule
        Model & Resolution & $|B_{\mathrm{eval}}|$ & $B_{\mathrm{eval}}$ (Q/DQ operator index) \\
        \midrule
        MobileNetV4 & $224\times224$ & 4 &
        indices 22, 25, 28, 31 \\
        MobileNetV4 & $512\times512$ & 4 &
        indices 22, 25, 28, 31 \\
        MobileNetV4 & $768\times768$ & 4 &
        indices 22, 25, 28, 31 \\
        MobileNetV4 & $1024\times1024$ & 4 &
        indices 22, 25, 28, 31 \\
        \midrule
        Hiera-Tiny & $224\times224$ & 2 &
        indices 1, 3 \\
        Hiera-Tiny & $384\times384$ & 2 &
        indices 1, 3 \\
        \bottomrule
    \end{tabular}
\end{table*}

\subsection{Selected boundary sets}
\label{sec:suppl_boundary_sets_embedded}

Following Eq.~\eqref{eq:feasible_set}, we tested every graph edge and used
the compatibility-screened pool to construct a jointly compilable
$B_{\mathrm{eval}}$, subject to $|B_{\mathrm{eval}}|\le K=4$, without
applying the deployment threshold. Candidates that failed the combined
compilation check were skipped as in Algorithm~\ref{alg:minimal_split_alg}.
Table~\ref{tab:s_boundary_set} reports the resulting evaluation sets used
in Table~\ref{tab:main_results}.
Detailed layer configurations appear in
Table~\ref{tab:s_mnv4_layers} for MobileNetV4 and
Table~\ref{tab:s_hiera_layers} for Hiera-Tiny.

\paragraph{Indexing convention.}
The reported indices refer to Q/DQ barrier operators after rewriting,
not to indices in the original graph. For MobileNetV4, Q/DQ barriers are
placed at indices 22, 25,
28, and 31, while for Hiera-Tiny they are at indices 1 and 3. Empirically,
all tested resolutions of a given model yielded the same selected
locations. See Section~\ref{subsec:safety_layer} for the feasibility test.

\subsection{Operator-level layer listings}

\begin{table*}[!t]
    \centering
    \footnotesize
    \caption{Layer configuration of MobileNetV4\_Hybrid\_Large. Indices are 0-based in operator order. Quant--dequant barriers mark split candidates.}
    \label{tab:s_mnv4_layers}
    \begin{tabular}{rllp{4.5cm}c}
        \toprule
        Idx & Type & Class & Description & Split \\
        \midrule
        0  & op      & conv2d                        & channels=$3\rightarrow 24$           &            \\
        1  & op      & batch\_norm\_act              &                               &            \\
        2  & pattern & edge\_residual                & channels=$24\rightarrow 48$               &            \\
        3  & pattern & universal\_inverted\_residual & channels=$48\rightarrow 96$               &            \\
        4  & pattern & universal\_inverted\_residual & channels=$96\rightarrow 96$               &            \\
        5  & pattern & universal\_inverted\_residual & channels=$96\rightarrow 192$              &            \\
        6  & pattern & universal\_inverted\_residual & channels=$192\rightarrow 192$             &            \\
        7  & pattern & universal\_inverted\_residual & channels=$192\rightarrow 192$             &            \\
        8  & pattern & universal\_inverted\_residual & channels=$192\rightarrow 192$             &            \\
        9  & pattern & universal\_inverted\_residual & channels=$192\rightarrow 192$             &            \\
        10 & pattern & universal\_inverted\_residual & channels=$192\rightarrow 192$             &            \\
        11 & pattern & universal\_inverted\_residual & channels=$192\rightarrow 192$             &            \\
        12 & pattern & mqa                           &                               &            \\
        13 & pattern & universal\_inverted\_residual & channels=$192\rightarrow 192$             &            \\
        14 & pattern & mqa                           &                               &            \\
        15 & pattern & universal\_inverted\_residual & channels=$192\rightarrow 192$             &            \\
        16 & pattern & mqa                           &                               &            \\
        17 & pattern & universal\_inverted\_residual & channels=$192\rightarrow 192$             &            \\
        18 & pattern & mqa                           &                               &            \\
        19 & pattern & universal\_inverted\_residual & channels=$192\rightarrow 192$             &            \\
        20 & pattern & universal\_inverted\_residual & channels=$192\rightarrow 512$             &            \\
        21 & pattern & universal\_inverted\_residual & channels=$512\rightarrow 512$             &            \\
        22 & op      & quant\_dequant\_barrier       & Q/DQ barrier                  & \checkmark \\
        23 & pattern & universal\_inverted\_residual & channels=$512\rightarrow 512$             &            \\
        24 & pattern & universal\_inverted\_residual & channels=$512\rightarrow 512$             &            \\
        25 & op      & quant\_dequant\_barrier       & Q/DQ barrier                  & \checkmark \\
        26 & pattern & universal\_inverted\_residual & channels=$512\rightarrow 512$             &            \\
        27 & pattern & universal\_inverted\_residual & channels=$512\rightarrow 512$             &            \\
        28 & op      & quant\_dequant\_barrier       & Q/DQ barrier                  & \checkmark \\
        29 & pattern & universal\_inverted\_residual & channels=$512\rightarrow 512$             &            \\
        30 & pattern & universal\_inverted\_residual & channels=$512\rightarrow 512$             &            \\
        31 & op      & quant\_dequant\_barrier       & Q/DQ barrier                  & \checkmark \\
        32 & pattern & universal\_inverted\_residual & channels=$512\rightarrow 512$             &            \\
        33 & pattern & universal\_inverted\_residual & channels=$512\rightarrow 512$             &            \\
        34 & pattern & mqa                           &                               &            \\
        35 & pattern & universal\_inverted\_residual & channels=$512\rightarrow 512$             &            \\
        36 & pattern & mqa                           &                               &            \\
        37 & pattern & universal\_inverted\_residual & channels=$512\rightarrow 512$             &            \\
        38 & pattern & mqa                           &                               &            \\
        39 & pattern & universal\_inverted\_residual & channels=$512\rightarrow 512$             &            \\
        40 & pattern & mqa                           &                               &            \\
        41 & pattern & universal\_inverted\_residual & channels=$512\rightarrow 512$             &            \\
        42 & pattern & conv\_bn\_act                 &                               &            \\
        43 & pattern & classification\_head          &                               &            \\
        \bottomrule
    \end{tabular}
\end{table*}

\begin{table*}[!t]
    \centering
    \footnotesize
    \caption{Layer configuration of Hiera Tiny.
    Indices are 0-based in operator order. Quant--dequant barriers mark split boundaries.}
    \label{tab:s_hiera_layers}
    \begin{tabular}{rllp{4.5cm}c}
        \toprule
        Idx & Type & Class & Description & Split \\
        \midrule
         0 & pattern & hiera\_patch\_embed &
         patch embedding&  \\
         1 & op      & quant\_dequant\_barrier &
         Q/DQ barrier & \checkmark \\
         2 & pattern & hiera\_block &
         dim=96$\rightarrow$96, tokens=3136 &  \\
         3 & op      & quant\_dequant\_barrier &
         Q/DQ barrier & \checkmark \\
         4 & pattern & hiera\_block &
         dim=96$\rightarrow$192, tokens=3136 &  \\
         5 & pattern & hiera\_block &
         dim=192$\rightarrow$192, tokens=196 &  \\
         6 & pattern & hiera\_block &
         dim=192$\rightarrow$384, tokens=196 &  \\
         7 & pattern & hiera\_block &
         dim=384$\rightarrow$384, tokens=49 &  \\
         8 & pattern & hiera\_block &
         dim=384$\rightarrow$384, tokens=49 &  \\
         9 & pattern & hiera\_block &
         dim=384$\rightarrow$384, tokens=49 &  \\
        10 & pattern & hiera\_block &
         dim=384$\rightarrow$384, tokens=49 &  \\
        11 & pattern & hiera\_block &
         dim=384$\rightarrow$384, tokens=49 &  \\
        12 & pattern & hiera\_block &
         dim=384$\rightarrow$384, tokens=49 &  \\
        13 & pattern & hiera\_block &
         dim=384$\rightarrow$768, tokens=49 &  \\
        14 & pattern & hiera\_block &
         dim=768$\rightarrow$768, tokens=49 &  \\
        15 & pattern & hiera\_block &
         dim=768$\rightarrow$768, tokens=49 &  \\
        16 & pattern & hiera\_block &
         dim=768$\rightarrow$768, tokens=49 &  \\
        17 & op      & layer\_norm &
         layer normalization &  \\
        18 & op      & reshape &
         reshape to sequence &  \\
        19 & op      & adaptive\_avg\_pool\_1d &
         global average pooling (1D) &  \\
        20 & op      & flatten &
         flatten &  \\
        21 & op      & fc &
         fully-connected classifier &  \\
        \bottomrule
    \end{tabular}
\end{table*}
 
\end{document}